\documentclass[conference]{IEEEtran}
\IEEEoverridecommandlockouts
\usepackage[utf8]{inputenc}
\usepackage{cite}
\usepackage{amsmath,amssymb,amsfonts}
\usepackage{algorithmic}
\usepackage{graphicx}
\usepackage{textcomp}
\usepackage{xcolor}
\usepackage{subcaption}
\usepackage{float}
\def\BibTeX{{\rm B\kern-.05em{\sc i\kern-.025em b}\kern-.08em
    T\kern-.1667em\lower.7ex\hbox{E}\kern-.125emX}}
\begin{document}

\title{LS-CAT: A Large-Scale CUDA AutoTuning Dataset}

\author{
\IEEEauthorblockN{Lars Bjertnes, Jacob O. Tørring, Anne C. Elster
}
\IEEEauthorblockA{\textit{Department of Computer Science} \\
\textit{Norwegian University of Science and Technology (NTNU)}\\
Trondheim, Norway \\
larbje@stud.ntnu.no, jacob.torring@ntnu.no, elster@ntnu.no
}
}

\maketitle

\begin{abstract}
The effectiveness of Machine Learning (ML) methods depend on access to large suitable datasets. In this article, we present how we build the LS-CAT (Large-Scale CUDA AutoTuning) dataset sourced from GitHub for the purpose of training NLP-based ML models. Our dataset includes 19 683 CUDA kernels focused on linear algebra. In addition to the CUDA codes, our LS-CAT dataset contains 5 028 536 associated runtimes, with different combinations of kernels, block sizes and matrix sizes. The runtime are GPU benchmarks on both Nvidia GTX 980 and Nvidia T4 systems. This information creates a foundation upon which NLP-based models can find correlations between source-code features and optimal choice of thread block sizes.

There are several results that can be drawn out of our LS-CAT database. E.g., our experimental results show that an optimal choice in thread block size can gain an average of 6\% for the average case. We thus also analyze how much performance increase can be achieved in general, finding that in 10\% of the cases more than 20\% performance increase can be achieved by using the optimal block. A description of current and future work is also included.

\end{abstract}


\section{Introduction}
To get full system usage, software implementations have to target each system. However, With the increase in hardware variations, the interface between software and hardware is becoming more complex. However, by parameterizing the software, this interface could be configurable. By tuning the parameters, the software could more efficiently interact with the hardware, resulting in increased performance. The challenge is that there are a lot of different parameters, that each usually have certain specific limits and legal values. Setting good parameters for a program requires high competence, while also being a time consuming process.

Autotuning attempts to solve this problem by computerizing the parameter adjustments. The system adjusts each parameter to some extent, then compiles, executes and measures the program. By comparing different results, the optimal combination of parameters can be found. 

There is usually a complex relation between a specific parameter and the total change in performance. A naive search through the parameter space is therefore typically not optimal. Since searching through all possible legal combinations of parameters is also a highly time consuming process, some methods have been developed \cite{b1,b2,b15} to search through the parameters more efficiently.

However, these methods are still reliant on compiling and executing the program to do gradual adjustments, which still takes a lot of time. A better alternative would be to have an autotuner that can find good parameters without compiling or executing the program.  

A dataset consisting of the results for each legal combination of hardware systems, and all other information that can change the performance or the run-time, could be used to find the absolute optimal combination of parameters for any given configuration. Unfortunately creating such a dataset would, as mentioned above, take incredibly long time, and the amount of data required would make the dataset huge. However, one could create a smaller dataset, that has a good enough coverage of the parameter space. This dataset could be used to find good parameters without compiling and executing everything, if there were a method to use this smaller dataset that is almost as good as having the entire dataset. 
Machine Learning (ML) is well suited in situations where there is enough data, and there is an unclear relationship between the start and end point. There are several machine learning models, but they all conceptually create an internal mathematical model typically consisting of activation's, weights, and biases. The weights and biases are adjusted depending on how the models output value compares to the target value. The model "learns" patterns this way, and can be used in a wide range of applications.

ML-assisted autotuning implementations have mostly focused on only the parameters and hardware information, and mostly using source code meta features \cite{b3}. However, the entirety of the source code plays a role in the software and should be taken into consideration.

ML-based attempts at using the whole source code for autotuning include finding the run time of parallel programs \cite{b4}, device mapping \cite{b5}, or multiple tasks \cite{b6,b7,b8,b9}. These attempts have usually focused on the programming languages C, C++ and OpenCL. OpenCL is a programming language that makes it possible to utilize the GPU for more general purpose programming. 
CUDA is designed specifically for Nvidia GPUs. The upside is that CUDA has a higher performance compared to OpenCL \cite{b10}. 

The earlier attempts at using source code to do ML-based autotuning on OpenCL, while getting good results, have limited themselves to a low number of distinct source codes. A lot of the data is sourced from libraries, which might not be representative of most written code. In this paper, we will, however, present how we generated a larger CUDA dataset sourced from a collection of independent developers.

\section{Background}

\subsection{GPU and CUDA}
GPUs have high core count that can process a lot of data simultaneously. CUDA, which is C++ based targets Nvidia GPUs. CUDA functions that run on the GPU are called kernels, and are either marked as global if run from the host system, or device if called from the global kernel. The global kernels take blocks of threads issued over a grid. The block parameter is a three dimensional representation of a collection of threads. Each block should be divisible by 32, which is known as the warp size. A warp executes all threads simultaneously. A block can at most run 1024 threads at the same time. The optimal number of threads per block is not always 1024, as several smaller blocks would have more unhindered register access, for instance.   


\subsection{Machine learning and NLP} 
Machine learning relies on two important concepts, the forward and backward pass. The forward pass is done by iterating over mathematical functions using an input parameter. In supervised learning, the output is then compared with the target value for the input. This comparison is done using a loss function that tries to find an accurate number for the difference of all output and target values at that given training step. This loss gives the model an adjustment target. A backward pass is then done by adjusting its internal weights and biases.

By repeating the process of forward and backward passes, the weights are adjusted to minimize the loss function. This, in turn, achieves outputs with similar values to the target values. 

As datatypes fed into a machine learning model have to be represented numerically, source code can't without any processing be fed directly into the model.  

Natural language processing (NLP), is focused on how to best represent text as numerical vectors. By using NLP techniques, source code can be transformed into distinct vectors, which can in turn be used for machine learning. 

\section{Related works}
\subsection{end2end-dl/deeptune}
Manually crafted heuristics are usually not ideal and can be challenging to create. Deeptune by Cummins et al. \cite{b6} therefore used an end-to-end ML-based model to create heuristics automatically. The dataset itself consists of a handful of unique OpenCL kernels, executed with different hardware systems on the GPU or CPU, and with varying number of threads. The source code is stored as raw code, but is pre-processed, discarding unnecessary information. Each line of code is then turned into a sequence of tokens, and turned into embeddings by the model. 


\subsection{NCC}
By using some of the methods from NLP, combined with code dependencies, NCC \cite{b8} tries to create an embedded representation of code based on LLVM IR or an intermediate representation of the code. 
Since it uses LLVM IR, the model should work on languages that can be compiled with LLVM. The embedder "inst2vec" can therefore be trained on a larger general purpose dataset, consisting of library sources. NCC then train on a smaller dataset that has a specific task, and the OpenCL dataset from DeepTune is reused for the same tasks. 




\subsection{CDFG}
CDFG \cite{b9} uses graph-based ML-techniques, unlike DeepTune and NCC, who focus on the sequenced part of source code learning. CDFG focuses on device mapping and thread coarsening using the same DeepTune and inst2vec datasets. 

One significant change made to the dataset is the addition of an abstract syntax tree (AST). The AST is a graph representation of how the code parts depend on each other. CDFG also labels all the connections so that they are not interchanged. 



\subsection{ProGraML}
ProGraML \cite{b7} further build upon using graph representations, by using three different graphs derived from source code. The goals are device mapping and algorithm classification, on the same DeepTune dataset. 

The three graphs used are control flow, data flow, and call flow. The control flow graph represents the order the statements are executed in based on their sequence and branching. The data flow graph is a data dependency graph. The call flow graph connects the original instruction jumps and the destinations, or the connection between called functions and where they were called from. This combined graph representation is made by using IR that has been normalized using inst2vec. 

ProGraML does not compare itself with CDFG, but with both DeepTune and NCC. Here ProGraML achieved the best results when compared with the others in device mapping, and algorithm classification.




\subsection{Public source code datasets and GH Archive}
To make our large structured source code dataset from GitHub we use GH Archive, which is a public archive of GitHub audit logs from the period of 2011 to now. Each log includes the JSON-encoded events reported by the GitHub API.
The entire GH Archive is also available as a public dataset on Google BigQuery, enabling
SQL searches on the dataset. 


\section{Data generation pipeline}

\subsection{Source code gathering}

The GH Archive dataset consist of several fields, the important ones in this case are the repository URL and payload. The payload contains meta data, and for instance the repository languages that have been used. To find which of the repositories that have a connection with CUDA or is CUDA related an SQL query to match the word CUDA can be used. In total there were 18534 unique repositories with the keyword CUDA.

There are several different ways to get the source code from GitHub. GitHub has multiple APIs that can be used by for instance Python. It is also possible to script the GitHub cloning function. The last way to retrieve the data is by downloading a repository as a zip file, which can also be automated by Python. Both methods were evaluated and timed, to find the faster one. Downloading as a zip using Python proved to be around twice as fast as cloning. 

The script that downloads the repositories creates a folder structure that corresponds to the indices in the repository URL file. 
Repositories from GitHub more often than not contains unneeded file types, by filtering out files based on file ending. Each repository were left with just C, C++, CUDA files and headers for the different languages.

The total amount of downloaded repositories were lower than the amount of URLs, which is explained by either the repositories being changed from public to private or that the repositories have been removed. In total 16247 repositories were downloaded successfully.

\subsection{Combining and executing code}

This section describes how we formatted and made the CUDA source code runnable.

To get any type of performance results from a repository, one might naively try to compile and execute the repository. This, however, is not a simple process at all -- especially if this process needs to be automated. The first part of compiling the repository is very dependent on a good Makefile or something similar, and even then any reliance on an external library would halt the automation significantly. 

Out of a hundred random repositories, only three managed to compile "out of the box". Even if all the repositories compiled flawlessly, an even bigger problem arises. How would one measure the performance, and the impact of modifications, on a random program? One might measure the entire program execution, or each CUDA kernel, but this would require to know specific program characteristics especially if the program had any kind of user interference. 

The solution to both of these issues were to find, isolate, and build each CUDA global function. Then each CUDA kernel could be run on its own, and measured accurately, with different modifications. The scripting for this process was done in Python, due to its ease of interacting with folders, files, and text. This is a multi-step process which can be summarized as follows:
\begin{itemize}
  \item Identify all global and device functions in repository.
  \item For each global function, find all relevant includes and device functions.
  \item Store global function as new file in new sub-folder.
  \item For each include any iterative dependency is found and added to the sub-folder.
  \item Find, format and store all global input parameters as list of tuples with names and types.
  \item Lastly, a generator uses the parameter information and naming to create a main file that can initialize all the needed variables.
  \item This main file can then issue a call to the global function.  
\end{itemize}


\subsection{Sampling and restructuring}
Even though the kernels are built, they still need to run, and additional work is needed to identify working kernels. With extra measures taken to fix broken kernels. 
To find out if a kernel has been properly isolated and can also be executed, a script is responsible for restructuring and compiling all the isolated kernels. 
With Python sub-processes, Python can execute any Linux command, including using the NVCC compiler and executing the kernels. 

\begin{figure}[H]
    \centering
    \includegraphics[width=0.75\linewidth]{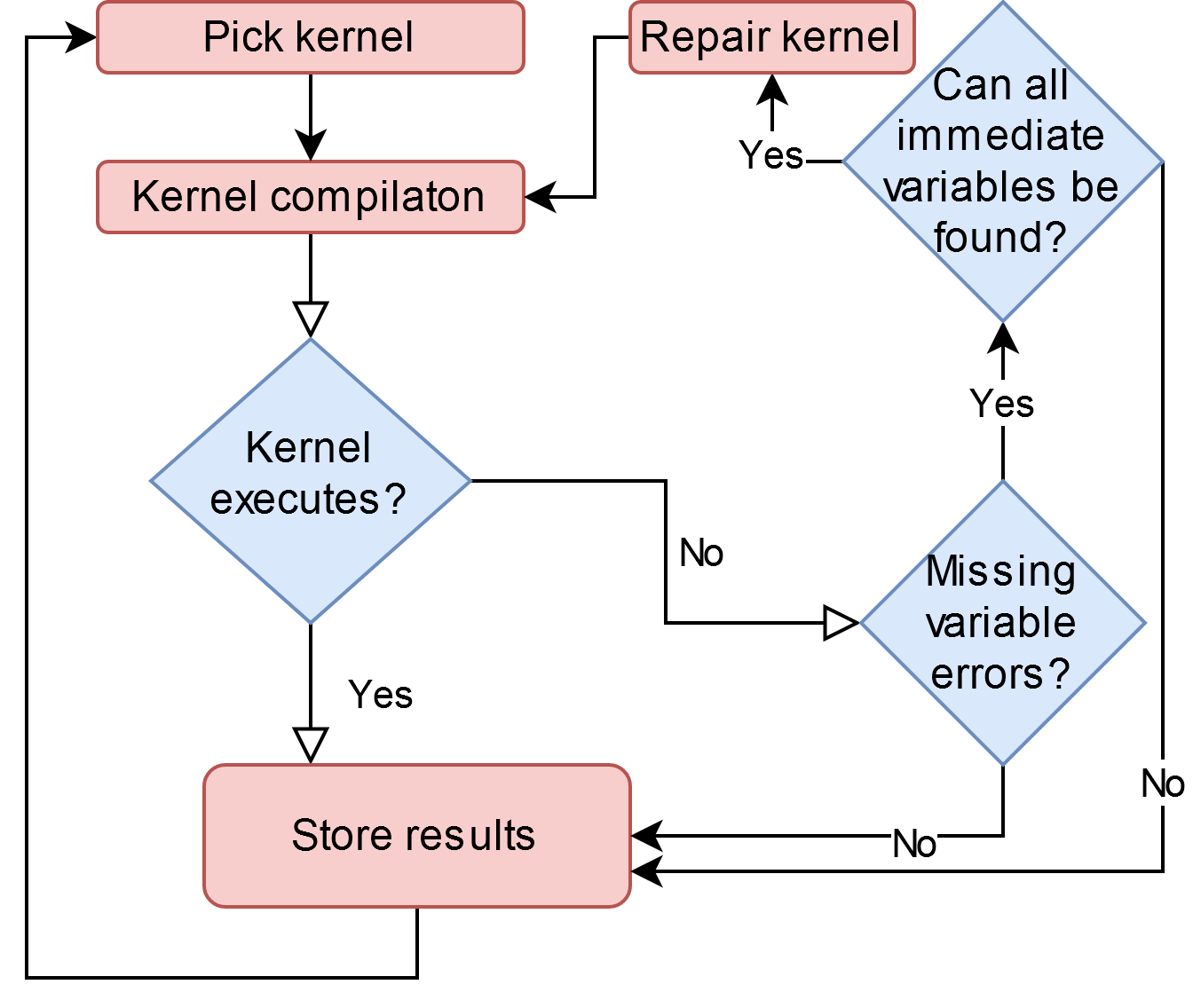}
    \caption{Restructuring and sampling}
    \label{fig1}
\end{figure}

This process of repeatedly fixing and compiling in Fig.~\ref{fig1} increased the amount of working kernels in the trial phase from around 22\% to 29\%.

After the time consuming process of compiling and fixing the potential kernels was done, 20 251 isolated kernels were compiling and could be executed as is. Additionally since all the errors were stored, potential improvements in the process of isolating the kernels could be identified by looking through the error messages. 

To get the results a new script was made which purpose was to modify the variables belonging to the kernel, compiling and executing, reading the outputted results and storing them. 

\subsection{Benchmarking}
Our experiments showed significant variations in run time for the same program. These were likely caused by external uncontrollable hardware and system processes. Bad results could make a machine learning model learn the wrong features. To mitigate some of these discrepancies, the official NVIDIA benchmark guide was followed~\cite{b12}.    

This guide is divided into two parts, locking hardware GPU and CPU clock speeds, and preheating the GPU. To preheat the GPU the program has to be executed once before any measurement is done on the actual program execution. This preheating ensures that the GPU spends minimal time going from an idle to active state, as this transition would impact run time. Similarly locking the hardware clock speed ensures that each program execution is stable without any boosting. Both of these changes decreases performance, but increases the accuracy of run time readings. On top of this the kernel itself is executed 1000 times, to minimize the variation of single run times. 

The NVIDIA guide improved the accuracy slightly, but there would still be some run time outliers large enough to impact the coherency of the result data. One way of tackling this is running the program several times, and aggregating the result again, in a way that makes the outputted result the most stable. Five different simple methods were tested out, by creating a dataset of 100 000 run times, then randomly selecting ten and aggregating, by repeating this process the variations after aggregation can be measured. Out of the different aggregation methods tested the median performed the best with a variation of around one percent.

\section{Experimental Setup}
We did initial tests of our dataset using a desktop Nvidia GeForce GTX 980 card based on Nvidia´s Maxwell architecture with 2048 CUDA Cores and 4GB of memory. We also used a newer system with 20 Nvidia T4s based on Nvidia´s Turing architecture with 2560 cores and 16GB memory each.

\subsection{Choice of parameters}
Of all the different parameters that could be tested the most reasonable ones to test, were matrix sizes and thread block sizes. This makes it also possible to compare with past OpenCL projects \cite{b6,b7,b8,b9}. 

To find parameter values that would be both reasonable to pick, and not too exhaustive as this would drastically increase the run-time, both the CUDA guidelines and similar projects were taken into consideration. 

The official CUDA guidelines, suggests using thread block sizes divisible by 32 and as large as possible. The largest block is 1024, and all blocks should be 32 divisible, leaving 32 different block sizes to be tested out, which is quite high.

Lim et al. \cite{b13} used the blocks in range of 0-1024 which were divisible by 64, this should be enough to achieve good results while also halving the amount of search space required. Additionally the 2D blocks of sizes (8,8) (16,16), (24,24) and (32,32) were also tested. 
To find a varying set of matrix dimensions some inspiration was taken from the Intel guide "measure the performance of matrix multiplications by dimensions" \cite{b14}.
In the end there were seven matrix sizes and twenty thread block sizes for a total of 140 combinations. 

\subsection{Kernel executor}
The last step is the execution of all the runnable kernels, with all the configurations needed. As the kernels were filtered into the ones that could execute and those that could not, the next step was modifying and setting variables, so that results could be extracted for every parameter combination.

The variable values were set mainly by their name, as the name-space proved to be a significant indicator for what kind of values they were supposed to take, w for width for instance or n for total size. The variables are split into either being set to the width, height, size, or 1 as some variables such as stride should remain static and within a reasonable range. Most, around 90\% of the variables, were covered by either width, height, or size or any other common name-space used in scientific applications, like k, the rest were also just set to one. 

The script decides first what type of template to use, and compiles the kernel using that template. After compilation the output executable can be executed with the Python sub-process, and each result is stored in a Pandas dataframe with the parameter combination, path, function name and result.

An important note is that when using the Python sub-process module the Linux timeout functionality should be used. The original reason it was implemented was to timeout kernel executions that were "hanging" and taking significantly longer time than other kernels. This does in effect filters out some of the kernels that might take too long to execute, and a potential ratio of how many kernel runs can be achieved vs the total time cost. For two seconds of timeout around one third of the kernels had enough time to execute, and at 120 seconds timeout only 1 in 500 did not execute, as a lower reasonable value, 30 seconds of timeout was used for the final multi GPU run. While two seconds as a test, were used on the GTX 980 system. A full execution on the GTX 980 took approximately 97 hours.

The cudaOccupancyMaxPotentialBlockSize is Nvidias own solution to find the optimal thread block size for a kernel, and was tested to see if it did indeed find the best size. These results could also be used in comparison with the final product, to do a comparative evaluation of a potential machine learnt model and this API. 
The cudaOccupancyMaxPotentialBlockSize API took significant less amount of time to execute, but was stopped halfway as the results indicated no result difference across matrix size.

The next step was executing this script with the same source data on our Nvidia Tesla T4 system. The T4 system has 20 GPUs. By using cudaSetDevice, the additional GPUs could run the script in parallel.  

 
\section{Results}
\subsection{GTX 980 System}
After the first full execution on the GTX 980 system we generated a dataset with 2140796 rows, and 97\% non NaN data.

\begin{figure}[]
    \centering
    \includegraphics[width=0.95\linewidth]{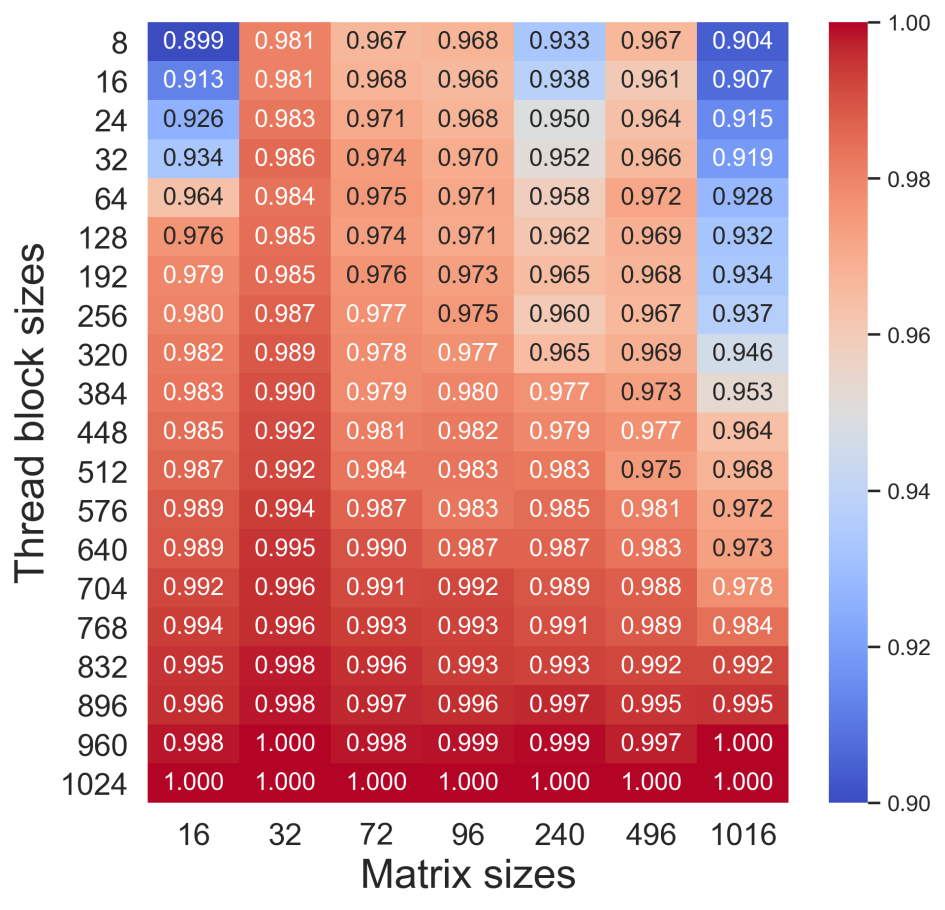}
    \caption{GTX 980 average performance of thread block sizes on matrix sizes}
    \label{fig2}
\end{figure}

We can see from Fig.~\ref{fig2} the downward trend in execution time caused by increase in thread block size, which both substantiate the claims by Nvidia regarding picking larger thread block sizes being the way to go, and that the results that were achieved were realistic. 

\begin{figure}[]
    \centering
    \includegraphics[width=0.95\linewidth]{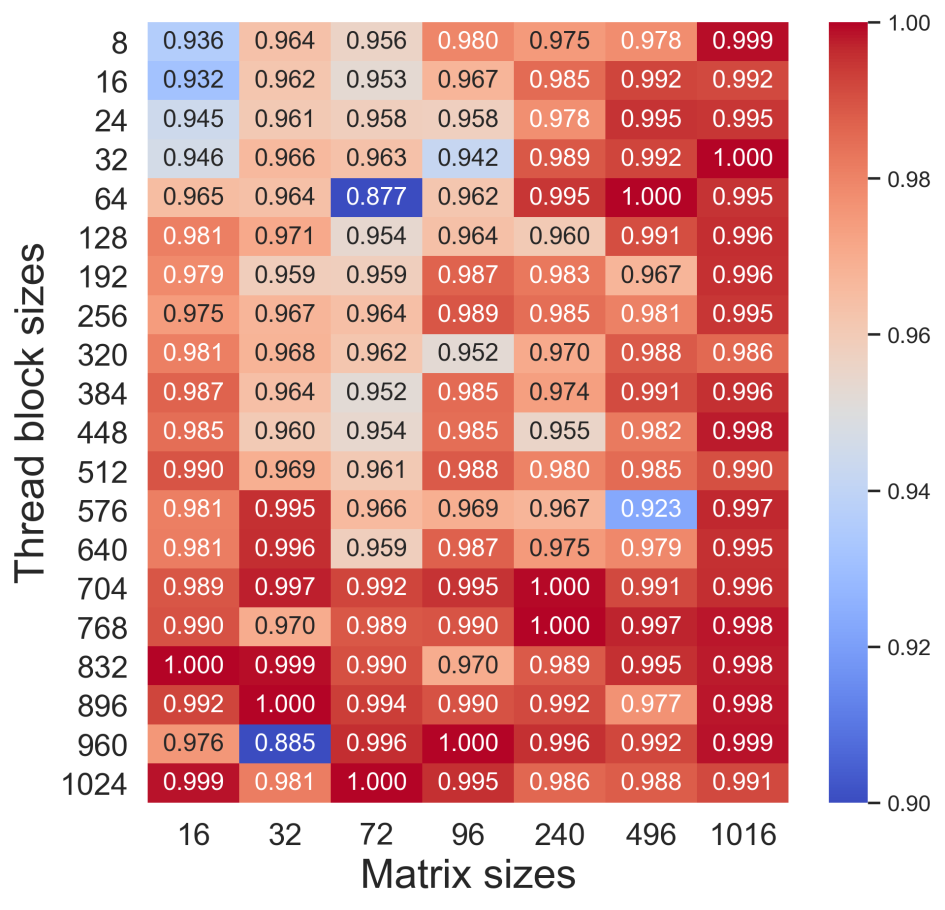}
    \caption{GTX 980 performance of thread block sizes on matrix sizes on the kernel euclidean\_kernel}
    \label{fig3}
\end{figure}

Now if the average was the best indicator and the variance between kernels thread block performance was low, then just picking a large thread block would always be the optimal choice. However if the graph of just one kernel is shown Fig.~\ref{fig3}, this no longer seems to be the case.

In this case the best performing thread block was not the largest one, and this was also true for 83\% of the kernels. The largest block did perform on average of 98.7\% of the best block, but in around 1\% of the kernels this ranges from 40 to 85\%, which would signify a large performance increase, if the better block size was used instead. 

\subsection{T4 System}
There was some small changes made to the matrix sizes, and an increase in timeout factor which did increase the amount of kernels with a run-time to 19683.

\begin{figure}[]
        \centering
        \includegraphics[width=0.95\linewidth]{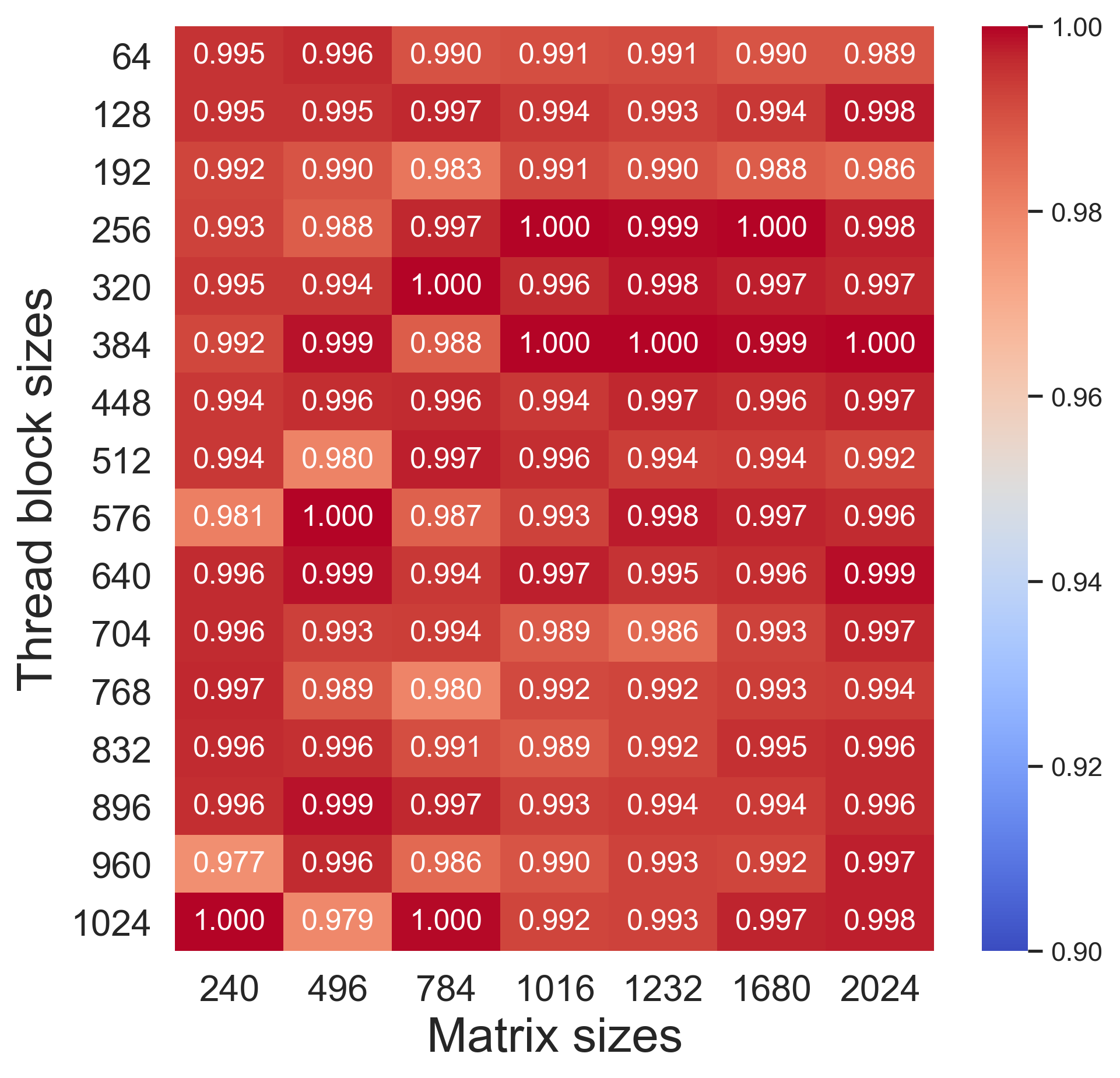}
        \caption{T4 Average performance of thread block sizes on matrix sizes}
        \label{fig4}
    \end{figure}

For the average kernel thread performance for each matrix size, there was quite a difference compared to GTX 980 Fig.~\ref{fig4}.

No longer is the average 1024 the best in all the average cases, which might either be due to increase in the amount of kernels which were recorded, or the change in hardware. 

\begin{figure}[]
    \centering
    \includegraphics[width=0.95\linewidth]{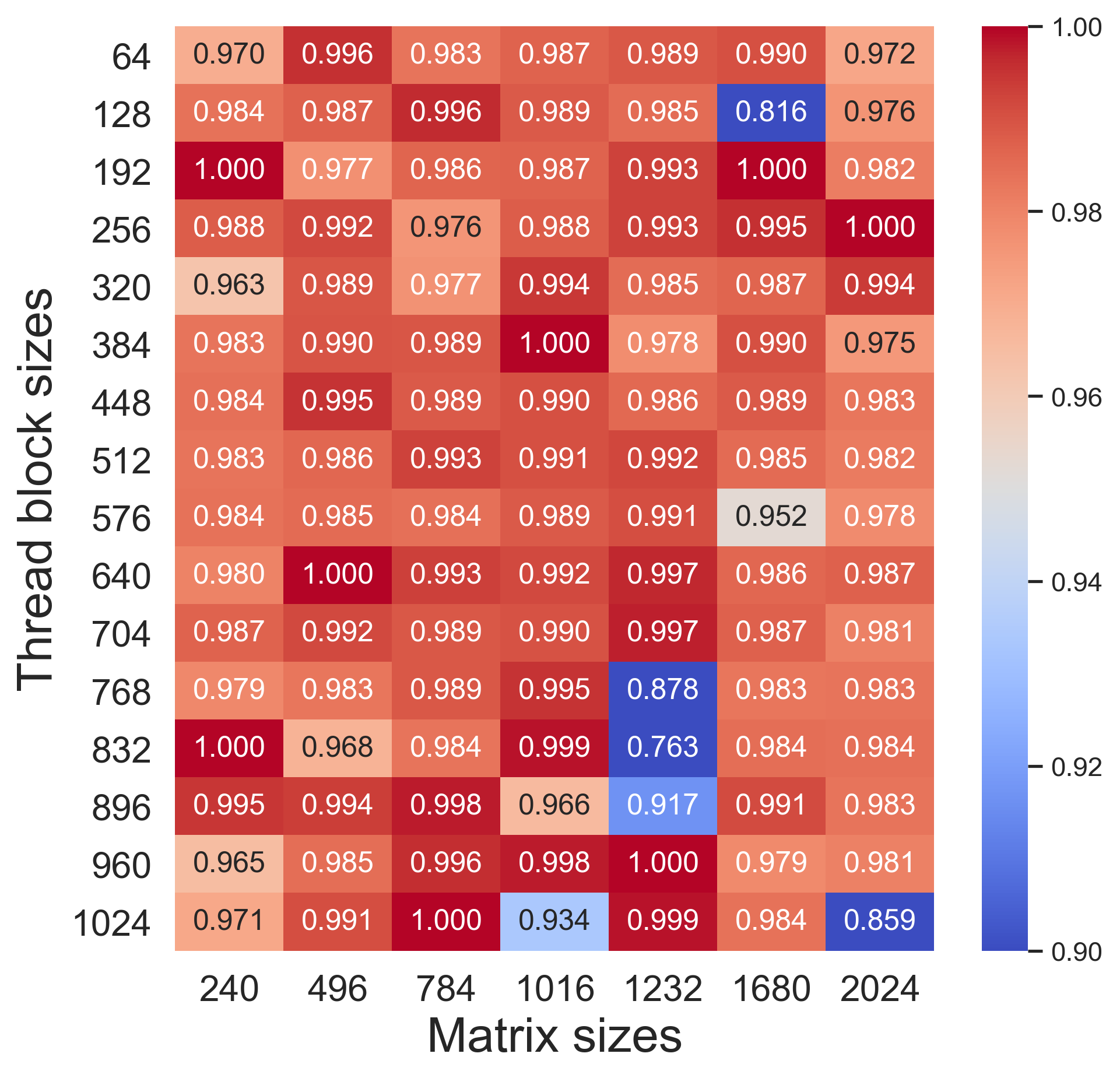}
    \caption{T4 performance of thread block sizes on matrix sizes on the kernel euclidean\_kernel}
    \label{fig5}
\end{figure}

Results from the same single kernel as in GTX 980, Fig.~\ref{fig5}.
    
It becomes quite evident that 1024 is not the perfect size in all cases, and a closer look shows that the 1024 block size has a 86\% performance compared to the best block. In 12\% of the cases a 1024 block would perform less than 85\% in comparison with the optimal block. 

\section{Discussion}
Compared to the other related works which are based upon Cummins et al.~\cite{b6} original dataset, our LS-CAT dataset is significantly larger. We present over 19 683 kernels compared to the Cummins et al. dataset of 256 kernels~\cite{b6}. The amount of variance in our dataset's programming style should also be impactful.

The range of different "topics" covered by the sheer difference in kernel amount would also be of help to more realistically evaluate the efficiency of machine learning on code, as unsupervised topic modeling could be used to see if some kernel topics are easier to evaluate than others.

To actually check the quality of the LS-CAT dataset for machine learning purposes, the dataset has to be used with a thread coarsening task in future works. If the machine learnt model performs well, the dataset is of sufficient quality. If the model performs poorly then either the dataset, or the methodology of using kernel code to find thread blocks is insufficient. 

If the LS-CAT dataset is proven to be insufficient, for whatever reason, there is enough modularity in the data processing pipeline and enough error information, from stored logs, to reconfigure the process to either increase data volume or quality. The public LS-CAT dataset will be updated in this case.

Another key finding was that 1024 proved to be a reasonable thread size for most cases, it might therefore be easier to identify the kernels for which this is not the case, than to find the best block for each kernel.

\section{Conclusions and Future Work}
In machine learning (ML), access to large enough datasets are very important for how well the resulting models perform. Previous work on using ML for autotuning CUDA codes, have had mixed results due to the lack of access to suitable code databases.

In this paper, we described how we generated a large-scale real-world dataset of CUDA kernels (LS-CAT\footnote{Available through https://www.ntnu.edu/idi/hpc-lab/}) for the purpose of training NLP-based ML-models for autotuning. 
The kernels were constructed by using source codes from GitHub via the GH Archive project~\cite{b11}.

We successfully downloaded 16 247 projects, out of the 18 534 projects that GH Archive showed as available, after pruning old and non-existent ones. Out of these, 20 251 runnable kernels were generated and compiled, and out of them again, 19 683 have results that we could use as a database of runable CUDA codes. 

In addition to the CUDA codes, our LS-CAT dataset contains 5 028 536 associated runtimes (including both GTX 980 and T4 results), with different combinations of kernels, block sizes and matrix sizes. 

Our experimental results coincided with what NVIDIA themselves have found, that increase in thread block size is usually enough, however, this is only true for the average case.

The results also indicate that always picking the optimal block over the largest, would net a 6\% performance increase on average, and in 10\% of the cases more than 20\% performance increase can be achieved by using the optimal block. Both of these findings are promising. 

The cudaOccupancyMaxPotentialBlockSize API was also tested to some extent, but proved insensitive to matrix sizes which does play a role in choice of blocks. 

Current and future work includes testing our LS-CAT dataset using NLP-ML models. 







\section*{Acknowledgment}
The authors would like to thank the Department of Computer Science and its HPC-Lab at NTNU for their support which enabled this project. The authors and the SFI Center for Geophysical Forecasting also plan to build on our LS-CAT dataset.


\end{document}